\documentclass{article}
\usepackage{amsmath}
\usepackage{graphicx}
\graphicspath{{images/}}
\usepackage[labelfont=bf]{caption}
\usepackage{float}
\usepackage{algpseudocode}
\usepackage{courier}
\usepackage{multicol}
\usepackage[procnames]{listings}
\newcommand{\pro}[1]{\operatorname{P}\left(#1\right)}
\newcommand{\len}[1]{\lvert#1\rvert}
\usepackage{dirtree}

\title{Predicting the large-scale evolution of tag systems}
\author{Carlos Martin}
\date{}

\begin{document}

\maketitle

\begin{abstract}
We present a method for predicting the large-scale evolution of a tag system from its production rules. A tag system's evolution is first divided into stages called `epochs' in which the tag system evolves monotonously. The distribution of symbols in the queue at the beginning of each epoch determines the tag system's large-scale properties, including growth rate and string densities, during that epoch. We derive the symbol distribution for the next epoch from the distribution for the current one, using this to make predictions over multiple successive epochs. Finally, we compare predictions that were obtained with this method to computer simulations and find that it retains great accuracy over several epochs.
\end{abstract}

\section{Introduction}

\subsection{Definition of a tag system}

A tag system is a model of computation comprised of a finite state machine and a queue. The queue contains symbols belonging to some alphabet \(\Sigma\). The finite state machine specifies a production function that maps strings of \(n\) symbols (elements of \(\Sigma^n\)) to strings of arbitrary length (elements of \(\Sigma^*\)).

In each step of a computation, \(n\) symbols are removed from the front of the queue and the corresponding string from the production function is added to the end of the same queue. This process is repeated until some halting condition is satisfied, such as there being fewer than \(n\) symbols in the queue.

Tag systems were created by the mathematician and logician Emil Leon Post, who is best known for his work in computability theory, as an example of a Post canonical system that is deterministic or \textit{monogenic}, meaning that at most one string can be produced from any given string in one step \cite{minsky1961}\cite{post1943}.

The emergence of complex behavior in very simple systems has been well-documented and explored in detail in Stephen Wolfram's \textit{A New Kind of Science} \cite{wolfram}. Wolfram's principle of computational equivalence states that almost all processes that are not obviously simple can be viewed as computations of equivalent sophistication. Even small systems can, in principle, compute the same things as any computer, given an appropriate translation of inputs and outputs.

It is therefore not entirely surprising that, despite the simplicity of their specification, tag systems have been shown to be capable of universal computation. This result was proven by Wang \cite{wang1963} and by Cocke and Minsky \cite{cocke1964} through the construction of a 2-tag system that can simulate a universal Turing machine. The search for other small universal tag systems remains open \cite{demol2008}\cite{rogozhin1996}.

Another interesting property of tag systems is their connection to problems in number theory. De Mol showed in \cite{demol2008} that the Collatz problem, a well-known unsolved problem in number theory, can be reduced to a small tag system. Furthermore, Conway proved in \cite{conway1972} that a generalization of the Collatz problem is algorithmically undecidable. Other undecidable variants of the Collatz problem are explored in \cite{lehtonen2008}. These examples support the principle of computational irreducibility described by Wolfram in \cite{wolfram}, which states that one cannot, in general, shortcut computations performed by sufficiently powerful automata.

Tag systems fall under this category. Although it is impossible to predict in general the exact behavior of tag systems, our study of their large-scale behavior could provide a useful heuristic to guide systematic searches for small tag system rules that can perform interesting computations, including ones connected to problems in number theory, like those of the Collatz problem.

Consider a simple 2-tag system with the following production rules:

\begin{equation}
aa \rightarrow aab,\,
ab \rightarrow ab,\,
ba \rightarrow b,\,
bb \rightarrow ba
\end{equation}

\begin{figure}[H]
\includegraphics[width=\textwidth]{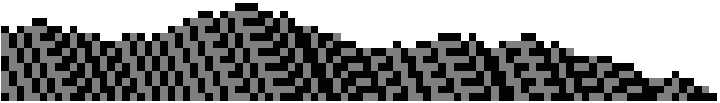}
\caption{}
\label{fig:small1}
\end{figure}

Figure \ref{fig:small1} illustrates the evolution of this tag system when initialized with a queue containing 10 symbols (where \(a\) symbols are colored gray and \(b\) symbols are colored black).

\begin{figure}[H]
\includegraphics[width=\textwidth]{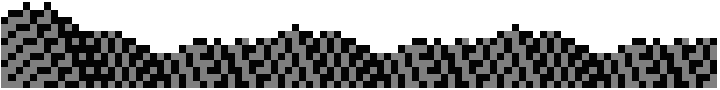}
\caption{}
\label{fig:small2}
\end{figure}

For this initial configuration, the tag system eventually reaches a state where it only has one symbol in the queue and thus terminates. Figure \ref{fig:small2} illustrates the same tag system under a different initial configuration. For this configuration, the tag system becomes periodic.

Now consider a tag system with a different set of production rules:

\begin{equation}
aa \rightarrow aba,\,
ab \rightarrow aa,\,
ba \rightarrow bbb,\,
bb \rightarrow ba
\end{equation}

\begin{figure}[H]
\includegraphics[width=\textwidth, height=100pt]{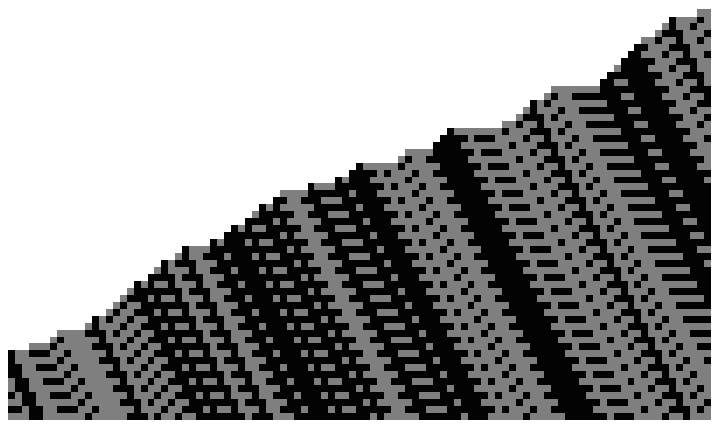}
\caption{}
\label{fig:small3}
\end{figure}

Figure \ref{fig:small3} illustrates the evolution of this tag system when initialized with a queue containing a random string of 10 symbols.

For this initial configuration, the length of the queue grows without bound. Furthermore, the queue of a tag system with these rules cannot contract because all productions have 2 or more symbols. Correspondingly, if all productions had 2 or fewer symbols, the queue of the tag system could never grow.

\subsection{Examples of large-scale evolution}

We can study the properties and behavior of tag systems like these on a much larger scale, revealing interesting statistical properties. The following images illustrate the evolution of various 2-tag systems for 5000 steps starting on a random initial state of 1000 symbols. Different random initial states all tend to produce roughly the same shape for each tag system, indicating that their large-scale behavior is determined primarily by their production rules.

\begin{figure}[H]
\includegraphics[width=\textwidth]{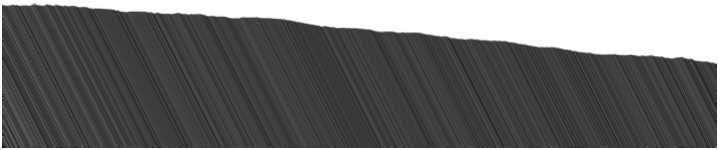}
\caption{\(
aa \rightarrow aab,\,
ab \rightarrow ab,\,
ba \rightarrow b,\,
bb \rightarrow ba
\)}
\label{fig:big1}
\end{figure}

\begin{figure}[H]
\includegraphics[width=\textwidth]{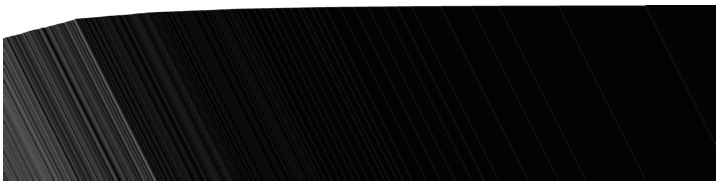}
\caption{\(
aa \rightarrow bb,\,
ab \rightarrow bb,\,
ba \rightarrow aaa,\,
bb \rightarrow bb
\)}
\label{fig:big2}
\end{figure}

The tag system shown in figure \ref{fig:big2} appears to initially grow at a rate of about 1 symbol every 4 steps, before it begins to level off and asymptotically approach a growth rate of zero. Furthermore, we can observe that the density of \(b\) symbols in the queue increases sharply beyond this point, while the density of \(a\) symbols decreases.

\begin{figure}[H]
\includegraphics[width=\textwidth, height=100pt]{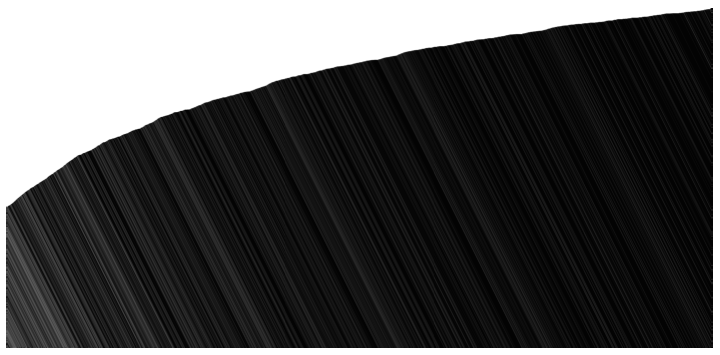}
\caption{\(
aa \rightarrow bab,\,
ab \rightarrow bbb,\,
ba \rightarrow aab,\,
bb \rightarrow bb
\)}
\label{fig:big3}
\end{figure}

\begin{figure}[H]
\includegraphics[width=\textwidth, height=100pt]{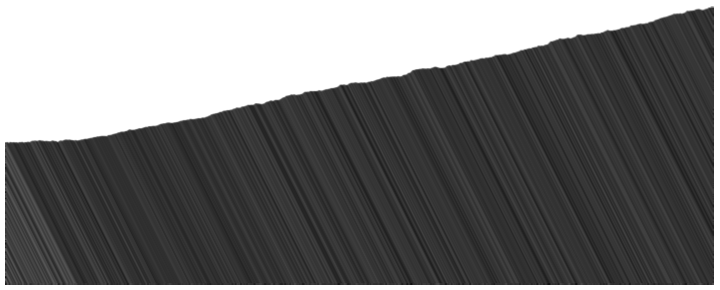}
\caption{\(
aa \rightarrow b,\,
ab \rightarrow b,\,
ba \rightarrow aab,\,
bb \rightarrow abb
\)}
\label{fig:big4}
\end{figure}

The tag system shown in figure \ref{fig:big4} is initially roughly constant in length, before transitioning to a long-term linear growth rate of approximately 1 symbol every 5 steps.

\begin{figure}[H]
\includegraphics[width=\textwidth]{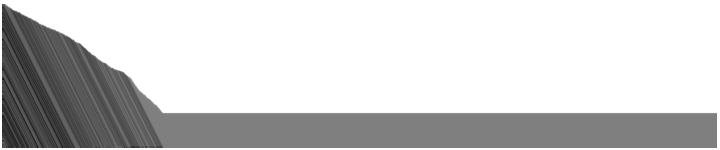}
\caption{\(
aa \rightarrow aa,\,
ab \rightarrow ba,\,
ba \rightarrow \varepsilon,\,
bb \rightarrow ab
\)}
\label{fig:big5}
\end{figure}

The tag system shown in figure \ref{fig:big5} has a phase of contraction before entering a phase where it remains constant in length and dominated by \(a\) symbols.

\begin{figure}[H]
\includegraphics[width=\textwidth]{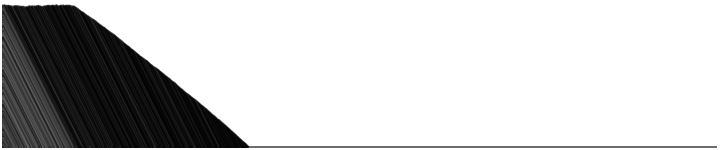}
\caption{\(
aa \rightarrow bbb,\,
ab \rightarrow ab,\,
ba \rightarrow bb,\,
bb \rightarrow b
\)}
\label{fig:big6}
\end{figure}

The tag system shown in figure \ref{fig:big6} enters a phase where it becomes entirely dominated by \(b\) symbols and remains constant in length, before entering another phase where it sharply contracts at a rate of approximately 1 symbol per step, until a single symbol remains.

\begin{figure}[H]
\includegraphics[width=\textwidth, height=100pt]{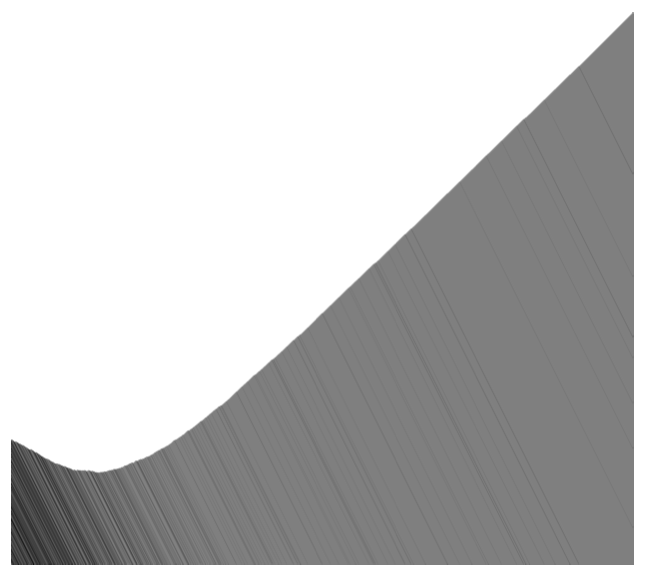}
\caption{\(
aa \rightarrow aaa,\,
ab \rightarrow b,\,
ba \rightarrow a,\,
bb \rightarrow b
\)}
\label{fig:big7}
\end{figure}

The tag system shown in figure \ref{fig:big7} at first contracts at a rate of approximately 1 symbol every 2 steps. It then transitions to a phase where it gradually starts approaching a growth rate of 1 symbol per step, while becoming increasingly dominated by \(a\) symbols. Its shape is reminiscent of an hourglass.

\begin{figure}[H]
\includegraphics[width=\textwidth]{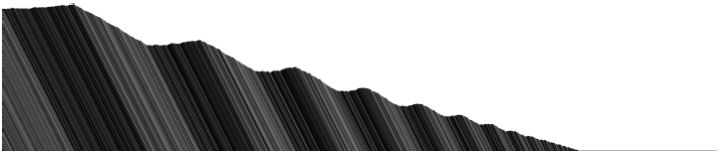}
\caption{\(
aa \rightarrow bbb,\,
ab \rightarrow ab,\,
ba \rightarrow bb,\,
bb \rightarrow a
\)}
\label{fig:big8}
\end{figure}

The tag system shown in figure \ref{fig:big8} exhibits a particularly interesting behavior. The tag system repeatedly alternates between two phases. In the first phase, the tag system becomes dominated by \(b\) symbols while remaining roughly constant in length. In the second phase, the densities of \(a\) and \(b\) symbols become roughly equal while the tag system contracts at a rate of approximately 1 symbol every 2 steps.

\subsection{Epochs and phase transitions}

There is a systematic way to study the large-scale growth patterns and behavior of these tag systems, even if they are non-linear (as demonstrated by some of the previous examples). We will use this approach to formulate an algorithmic procedure for predicting these large-scale properties.

The first insight needed to understand the large-scale evolution of these tag systems is that the history or evolution of the tag system can be divided naturally into distinct stages, which we will refer to as \textit{epochs}.

\begin{figure}[H]
\includegraphics[width=\textwidth]{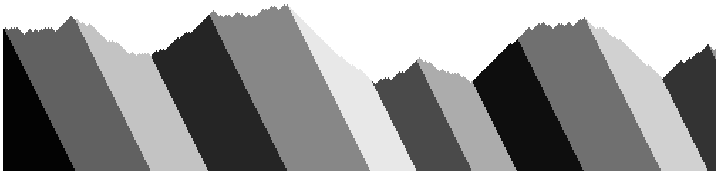}
\caption{}
\label{fig:big9}
\end{figure}

The beginning of a new epoch occurs when all the symbols belonging to the previous epoch have been consumed by the tag system (i.e. removed from the queue). Figure \ref{fig:big9} illustrates the evolution of a tag system where symbols in the queue are colored according to which epoch they were produced in.

\begin{figure}[H]
\includegraphics[width=\textwidth]{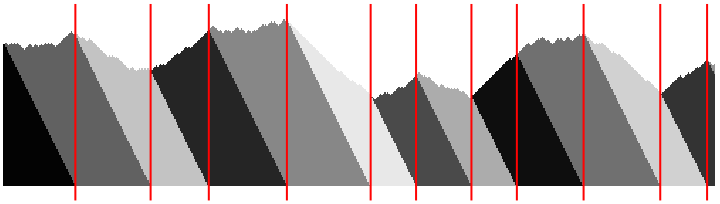}
\caption{}
\label{fig:big10}
\end{figure}

Figure \ref{fig:big10} illustrates the same evolution while highlighting the beginning of each epoch with a vertical line.

Notice that the length of the queue tends to change in a relatively linear manner within each epoch. The same can be said of the previous examples of 2-tag systems. Figure \ref{fig:big11} illustrates the evolution of the tag system with decelerating growth:

\begin{figure}[H]
\includegraphics[width=\textwidth, height=100pt]{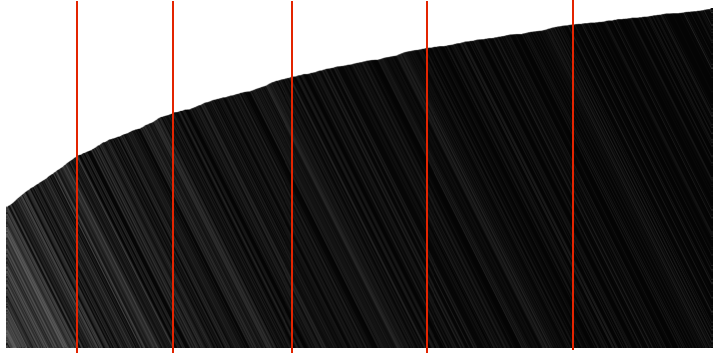}
\caption{}
\label{fig:big11}
\end{figure}

Figure \ref{fig:big12} illustrates the tag system which exhibits a strong phase transition from contraction to growth on the second epoch:

\begin{figure}[H]
\includegraphics[width=\textwidth, height=100pt]{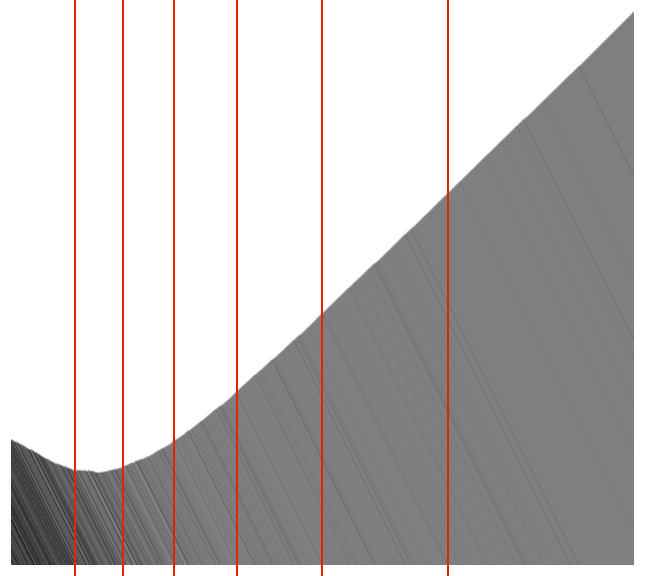}
\caption{}
\label{fig:big12}
\end{figure}

\begin{figure}[H]
\includegraphics[width=\textwidth]{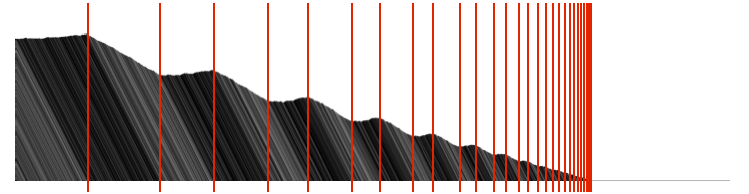}
\caption{}
\label{fig:big13}
\end{figure}

Figure \ref{fig:big13} illustrates the tag system which repeatedly transitions between a phase of constant length and a phase of constant contraction.

The remainder of this paper will make the assumption that the distributions of symbols and strings of symbols on the queue are stationary along its length, meaning they do not change significantly from the beginning to the end of the queue. If a distribution is stationary at the beginning of an epoch, it will likely remain stationary at the beginning of the next epoch, since the latter is generated from the former by the same set of production rules. The stationarity assumption means that the large-scale properties of a tag system tend to change linearly within epochs.

The second key insight needed to predict the large-scale evolution of a tag system is the following: The distribution of symbols on the queue for the beginning of an epoch is entirely a function of the distribution of symbols on the queue for the beginning of the previous epoch.

In particular, the distribution of \textit{tuples} of symbols on the queue for the beginning of an epoch can be used to determine the distribution of \textit{strings} of symbols for the next epoch (by analysing the production rules of the tag system). From this string distribution, one can determine the distribution of \textit{tuples} of symbols, which can, in turn, be used to determine the string distribution for the \textit{third} epoch, and so on. This procedure is outlined below:

\lstset{
    language=Python, 
    basicstyle=\ttfamily\small,
    showstringspaces=false,
    keywordstyle=\bfseries,
    commentstyle=\itshape
}

\begin{lstlisting}
word_probs[0] = initial_word_probs
for epoch in range(epochs):
  prod_probs[epoch + 1] = get_prod_probs(word_probs[epoch])
  word_probs[epoch + 1] = get_word_probs(prod_probs[epoch + 1])
\end{lstlisting}

where \texttt{word\_probs} is the distribution of \(n\)-tuples on the tag system queue and \texttt{prod\_probs} is the distribution indicating the probability of different string productions being generated at the beginning of an epoch.

In the next section, we will describe how to find the production distribution from the tuple distribution and, with more difficulty, the tuple distribution from the production distribution.

\section{Derivation}

\subsection{Generating a production}

Let \(p\) be the contents of the queue at the beginning of an epoch:

\begin{equation}
p = q_1 \cdot q_2 \cdot q_3 \cdot \ldots \cdot q_m
\end{equation}

where \(q_i \in \Sigma^n\) and \(\cdot\) denotes concatenation. The production rules constitute a function \(f\) that maps strings of length \(n\) to strings of arbitrary length:

\begin{equation}
f : \Sigma^n \rightarrow \Sigma^*
\end{equation}

\begin{figure}[H]
\includegraphics[width=\textwidth]{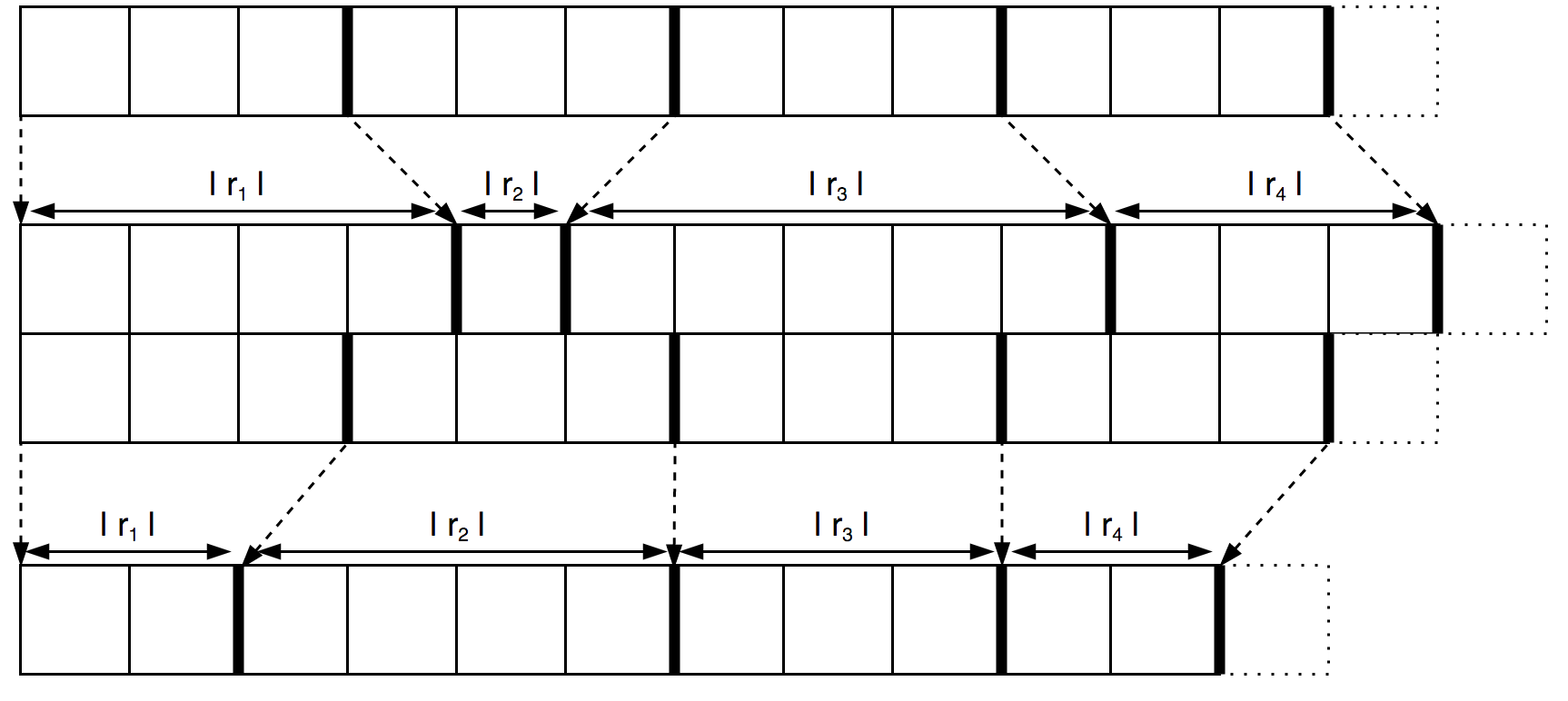}
\caption{}
\label{fig:diagram1}
\end{figure}

Hence the contents of the queue at the beginning of the next epoch are

\begin{equation}
\begin{split}
p &= f(q_1) \cdot f(q_2) \cdot f(q_3) \cdot \ldots \cdot f(q_m) \\
&= r_1 \cdot r_2 \cdot r_3 \cdot \ldots \cdot r_m
\end{split}
\end{equation}

where \(r_i \in \Sigma^*\). The concatenation of these productions, in turn, determines the productions that are generated in the epoch after the next one. This process is illustrated in figure \ref{fig:diagram1}.

The probability that \(r_i = s\) for some \(s \in \Sigma^*\) is

\begin{equation}
\begin{split}
\pro{r_i = s} &= \pro{f(q_i) = s} \\
&= \pro{\bigcup_{t \in \Sigma^n} q_i = t \cap f(q_i) = s} \\
&= \sum_{t \in \Sigma^n} \pro{q_i = t \cap f(q_i) = s} \\
&= \sum_{t \in \Sigma^n} \pro{q_i = t} \pro{f(q_i) = s \mid q_i = t} \\
&= \sum_{t \in \Sigma^n} \pro{q_i = t} \pro{f(t) = s} \\
&= \sum_{t \in \Sigma^n} \pro{q_i = t} [f(t) = s]
\end{split}
\end{equation}

since \(\pro{f(t) = s} \in \{0, 1\}\) if the tag system is deterministic. Consequently, given the distribution of strings of length \(n\) at the beginning of an epoch, one can determine the distribution of strings produced during that epoch. Because the queue at the beginning of the next epoch consists of the concatenation of these productions, one can theoretically determine the distribution of strings of length \(n\) at the beginning of the next epoch as well.

\subsection{Selecting a production instance}

Let \(i\) be a position selected uniformly at random from \(p\). Recall the definition of a uniform distribution:

\begin{equation}
X \sim \mathcal{U}(S) \Longleftrightarrow (\forall R \subset S) \left(\pro{X \in R} = \frac{\mu(R)}{\mu(S)}\right)
\end{equation}

where \(X\) is the random variable, \(S\) is the sample space, and \(\mu\) is a measure. In our case, \(X = i\) and \(S = [0, p)\):

\begin{equation}
\pro{i \in R} = \frac{\mu(R)}{\mu([0, p))}
\end{equation}

Let \(i \circ j\) be the statement that \(i\) belongs to \(r_j\):

\begin{equation}
\begin{split}
i \circ j &\Longleftrightarrow \len{r_1 \cdot \ldots \cdot r_{j-1}} \leq i < \len{r_1 \cdot \ldots \cdot r_{j-1} \cdot r_j} \\
&\Longleftrightarrow i \in [\len{r_1 \cdot \ldots \cdot r_{j-1}}, \len{r_1 \cdot \ldots \cdot r_{j-1} \cdot r_j})
\end{split}
\end{equation}

The probability of this is

\begin{equation}
\begin{split}
\pro{i \circ j} &= \pro{i \in [\len{r_1 \cdot \ldots \cdot r_{j-1}}, \len{r_1 \cdot \ldots \cdot r_{j-1} \cdot r_j})} \\
&= \frac{\mu([\len{r_1 \cdot \ldots \cdot r_{j-1}}, \len{r_1 \cdot \ldots \cdot r_{j-1} \cdot r_j}))}{\mu([0, p))} \\
&= \frac{\len{r_1 \cdot \ldots \cdot r_{j-1} \cdot r_j} - \len{r_1 \cdot \ldots \cdot r_{j-1}}}{\len{p}} \\
&= \frac{\len{r_j}}{\len{p}}
\end{split}
\end{equation}

Therefore, the probability of selecting a position that belongs to \(r_j\) is proportional to the length of \(r_j\). In other words, longer production instances are more likely to contain the position that was selected at random.

\begin{figure}[H]
\includegraphics[width=\textwidth]{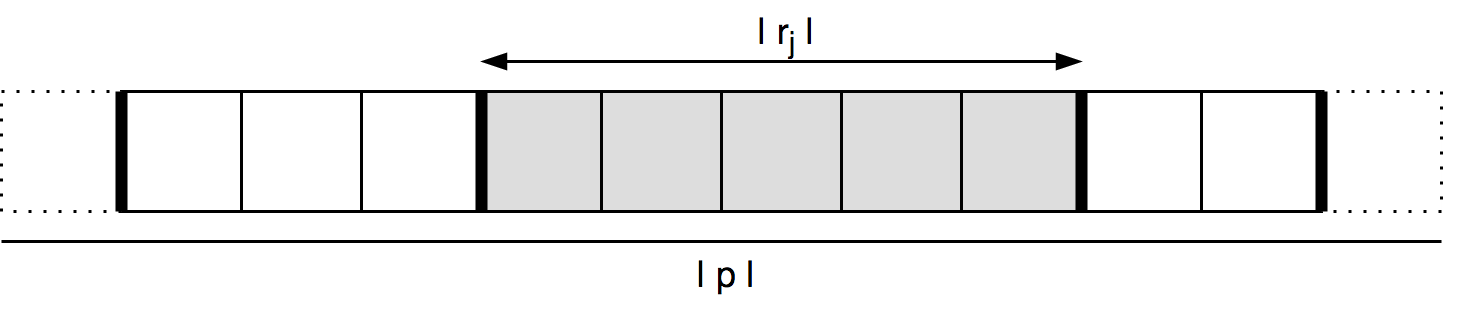}
\caption{}
\end{figure}

\subsection{Selecting a production type}

Let \(i \triangleleft s\) be the statement that \(i\) belongs to some \(r_j\) equal to \(s\):

\begin{equation}
i \triangleleft s \Longleftrightarrow \bigcup_{j \in [1, m]} i \circ j \cap r_j = s
\end{equation}

The probability of this is

\begin{equation}
\begin{split}
\pro{i \triangleleft s} &= \pro{\bigcup_{j \in [1, m]} i \circ j \cap r_j = s} \\
&= \sum_{j \in [1, m]} \pro{i \circ j \cap r_j = s} \\
&= \sum_{j \in [1, m]} \pro{r_j = s} \pro{i \circ j \mid r_j = s} \\
&= \sum_{j \in [1, m]} \pro{r_j = s} \frac{\len{s}}{\len{p}} \\
&= \frac{\len{s}}{\len{p}} \sum_{j \in [1, m]} \pro{r_j = s} \\
&= \pro{r_j = s} \frac{\len{s}}{\len{p}} \sum_{j \in [1, m]} 1 \\
&= \pro{r_j = s} \frac{\len{s}}{\len{p}} m \\
\end{split}
\end{equation}

\(\pro{r_j = s}\) is factored from the sum because it is independent of \(j\). Intuitively, the production type \(s\) is expected to appear \(\pro{r_j = s} m\) times in \(p\), where \(m\) is the total number of productions. Hence symbols belonging to \(s\) are expected to appear \(\pro{r_j = s} m \len{s}\) times. The probability that a randomly selected symbol belongs to \(s\) is the ratio of this quantity to the total number of symbols, which is the length of \(p\). The expected length of \(p\) can be found as follows:

\begin{equation}
\begin{split}
\operatorname{E}[\len{p}] &= \operatorname{E}[\len{r_1 \cdot r_2 \cdot r_3 \cdot \ldots}] \\
&= \operatorname{E}\left[\sum_{j \in [1, m]} \len{r_j}\right] \\
&= \sum_{j \in [1, m]} \operatorname{E}[\len{r_j}] \\
&= \operatorname{E}[\len{r_j}] \sum_{j \in [1, m]} 1 \\
&= \operatorname{E}[\len{r_j}] m \\
\end{split}
\end{equation}

where \(\operatorname{E}[\len{r_j}]\) is factored from the sum because it is independent of \(j\). Recall that the expected value of a function \(f\) of a random variable \(X\) is

\begin{equation}
\operatorname{E}[f(X)] = \sum_{x \in \Omega} \pro{X = x} f(x)
\end{equation}

where \(\Omega\) is the sample space. Hence

\begin{equation}
\operatorname{E}[\len{r_j}] = \sum_{s' \in \Sigma^*} \pro{r_j = s'} \len{s'}
\end{equation}

and the expected length of \(p\) is

\begin{equation}
\operatorname{E}[\len{p}] = m \sum_{s' \in \Sigma^*} \pro{r_j = s'} \len{s'}
\end{equation}

Therefore

\begin{equation}
\begin{split}
\pro{i \triangleleft s} &= \pro{r_j = s} \frac{\len{s}}{\len{p}} m \\
&= \frac{\pro{r_j = s} \len{s} m}{m \sum_{s' \in \Sigma^*} \pro{r_j = s'} \len{s'}} \\
&= \frac{\pro{r_j = s} \len{s}}{\sum_{s' \in \Sigma^*} \pro{r_j = s'} \len{s'}}
\end{split}
\end{equation}

In other words, the probability of selecting a position belonging to a production type is dependent on both the length of the production and the probability of that production being generated.

\subsection{Selecting a string given a production}

For any string \(s\), let \(s_{a : b}\) be the substring of \(s\) starting at position \(a\) (inclusive) and ending at position \(b\) (exclusive):

\begin{equation}
s_{a : b} = s_a s_{a+1} s_{a+2} \ldots s_{b-2} s_{b-1}
\end{equation}

For all \(r \in \Sigma^n\), one can determine the probability that \(q_{i : i + \len{r}} = r\) as follows:

\begin{equation}
\begin{split}
\pro{q_{i : i + \len{r}} = r} &= \pro{\bigcup_{s \in \Sigma^*} q_{i : i + \len{r}} = r \cap i \triangleleft s} \\
&= \sum_{s \in \Sigma^*} \pro{q_{i : i + \len{r}} = r \cap i \triangleleft s} \\
&= \sum_{s \in \Sigma^*} \pro{i \triangleleft s} \pro{q_{i : i + \len{r}} = r \mid i \triangleleft s}
\end{split}
\end{equation}

The conditional probability \(\pro{q_{i : i + \len{r}} = r \mid i \triangleleft s}\) depends on which position \(j\) in \(s\) has been selected, and can be determined as follows:

\begin{equation}
\begin{split}
\pro{q_{i : i + \len{r}} = r \mid i \triangleleft s} &= \pro{\bigcup_{j \in [0, \len{s})} q_{i : i + \len{r}} = r \cap j} \\
&= \sum_{j \in [0, \len{s})} \pro{q_{i : i + \len{r}} = r \cap j} \\
&= \sum_{j \in [0, \len{s})} \pro{j} \pro{q_{i : i + \len{r}} = r \mid j}
\end{split}
\end{equation}

Because \(i\) is selected uniformly at random from \(p\), \(j\) is also selected uniformly at random from \(s\):

\begin{equation}
\begin{split}
\pro{q_{i : i + \len{r}} = r \mid i \triangleleft s} &= \sum_{j \in [0, \len{s})} \pro{j} \pro{q_{i : i + \len{r}} = r \mid j} \\
 &= \sum_{j \in [0, \len{s})} \frac{1}{\len{s}} \pro{q_{i : i + \len{r}} = r \mid j} \\
 &= \frac{1}{\len{s}} \sum_{j \in [0, \len{s})} \pro{q_{i : i + \len{r}} = r \mid j}
\end{split}
\end{equation}

To determine \(\pro{q_{i : i + \len{r}} = r \mid j}\), one must first consider whether \(j + \len{r} < \len{s}\) or \(j + \len{r} \geq \len{s}\). If \(j + \len{r} < \len{s}\), one must check whether \(s_{j : j + \len{r}} = r\):

\begin{figure}[H]
\includegraphics[width=\textwidth]{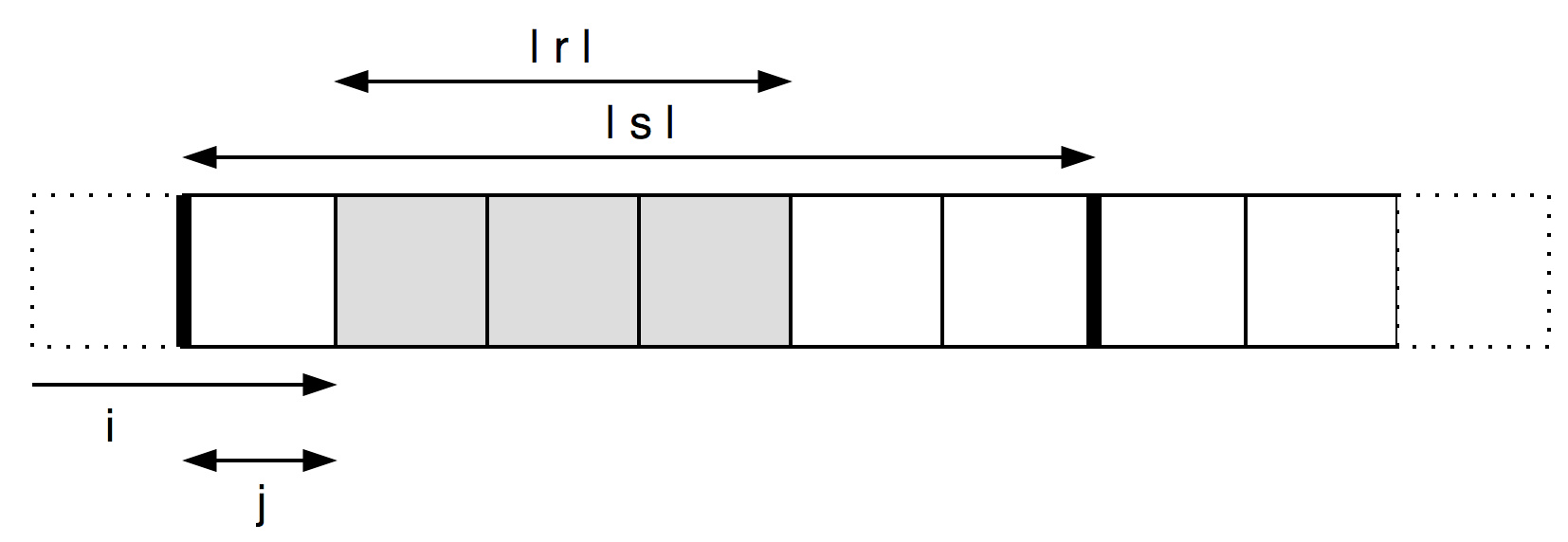}
\caption{}
\end{figure}

If \(j + \len{r} \geq \len{s}\), one must check whether \(s_{j : \len{s}} = r_{0 : \len{s} - j}\) \textit{and} whether the rest of \(p\) begins with the remaining substring of \(r\) (or \(p_{i - j + \len{s} : i + \len{r}} = r_{\len{s} - j : \len{r}}\)):

\begin{figure}[H]
\includegraphics[width=\textwidth]{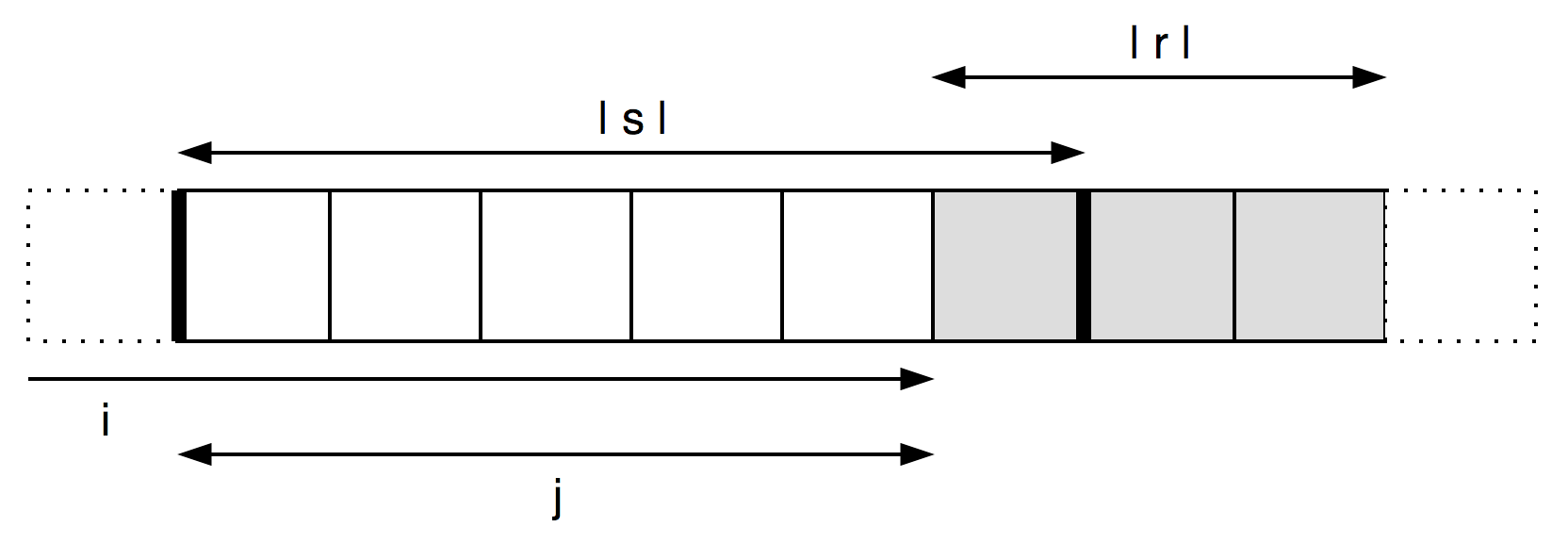}
\caption{}
\end{figure}

In summary:

\begin{equation}
p_{i : i + \len{r}} = r \Leftrightarrow \begin{cases}
s_{j : j + \len{r}} = r & j + \len{r} < \len{s} \\
s_{j : \len{s}} = r_{0 : \len{s} - j} \cap p_{i - j + \len{s} : i + \len{r}} = r_{\len{s} - j : \len{r}} & j + \len{r} \geq \len{s}
\end{cases}
\end{equation}

Therefore

\begin{equation}
\pro{p_{i : i + \len{r}} = r} \Leftrightarrow \begin{cases}
1 & j + \len{r} < \len{s} \cap s_{j : j + \len{r}} = r \\
\pro{p_{i - j + \len{s} : i + \len{r}} = r_{\len{s} - j : \len{r}}} & j + \len{r} \geq \len{s} \cap s_{j : \len{s}} = r_{0 : \len{s} - j}
\end{cases}
\end{equation}

\subsection{Production sequence beginning with a string}

Let \(p' = p_{i - j + \len{s} : \len{p}}\) denote the rest of \(p\), that is, the substring of \(p\) that consists of the concatenation of all the production instances following the production instance that contained the selected position \(i\):

\begin{equation}
p' = r_j \cdot r_{j+1} \cdot r_{j+2} \cdot \ldots
\end{equation}

The probability that \(p'\) begins with a string \(t\) is

\begin{equation}
\begin{split}
\pro{p'_{0 : \len{t}} = t} &= \pro{p'_{0 : \len{t}} = t \cap \bigcup_{s \in \Sigma^*} r_j = s} \\
&= \pro{\bigcup_{s \in \Sigma^*} p'_{0 : \len{t}} = t \cap r_j = s} \\
&= \sum_{s \in \Sigma^*} \pro{p'_{0 : \len{t}} = t \cap r_j = s} \\
&= \sum_{s \in \Sigma^*} \pro{r_j = s} \pro{p'_{0 : \len{t}} = t \mid r_j = s}
\end{split}
\end{equation}

\begin{figure}[H]
\includegraphics[width=\textwidth]{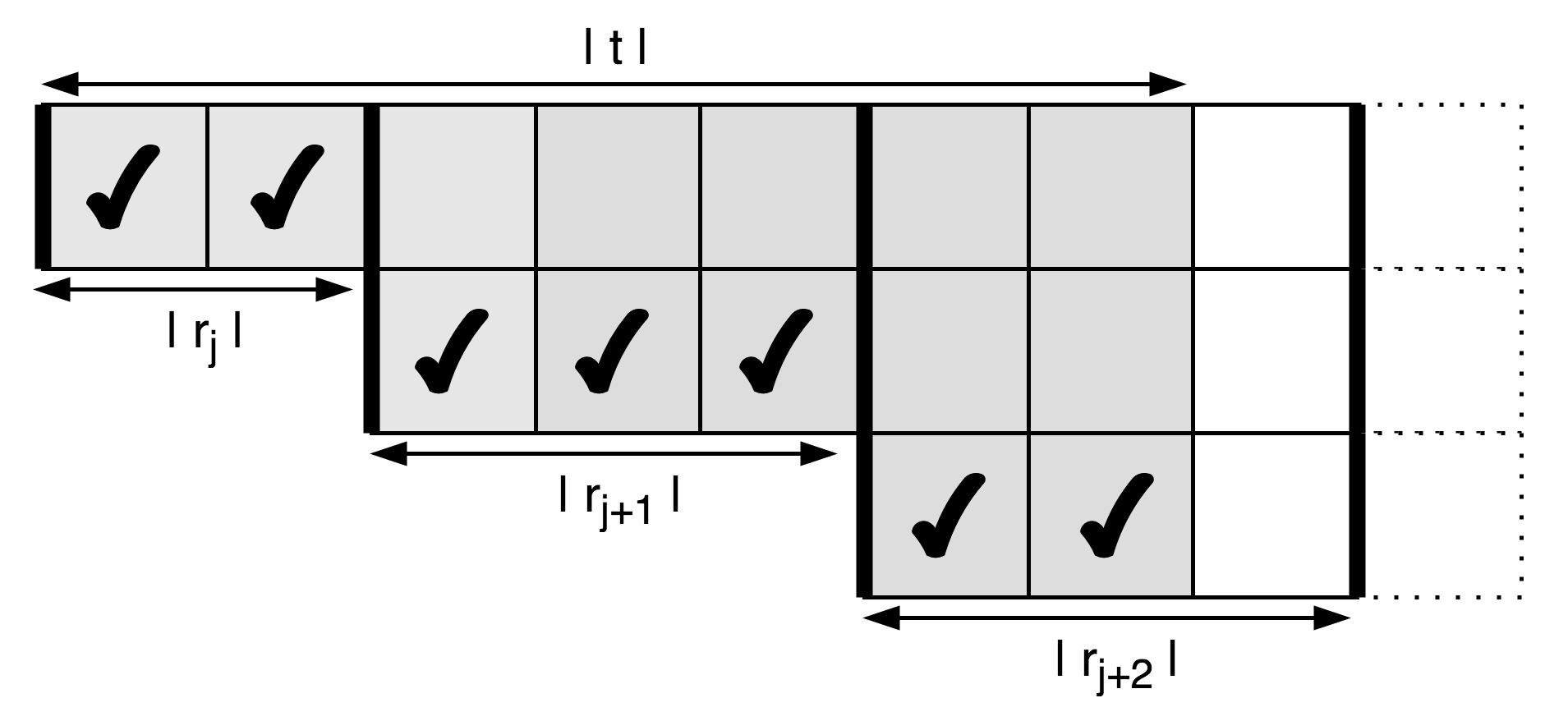}
\caption{}
\label{fig:diagram5}
\end{figure}

To determine \(\pro{p'_{0 : \len{t}} = t \mid r_j = s}\), one must first consider whether \(\len{t} \leq \len{s}\) or \(\len{t} > \len{s}\). If \(\len{t} \leq \len{s}\), then the sequence begins with \(t\) if the first \(\len{t}\) symbols of \(s\) match \(t\). Alternatively, if \(\len{t} > \len{s}\), then the sequence begins with \(t\) if the first \(\len{s}\) symbols of \(t\) match \(s\) \textit{and} the rest of the sequence begins with the remaining symbols of \(t\). This procedure is illustrated in figure \ref{fig:diagram5}.

This means the conditional probability can be expressed as

\begin{equation}
\pro{p'_{0 : \len{t}} = t \mid r_j = s} = \begin{cases}
1 & \len{t} \leq \len{s} \cap s_{0 : \len{t}} = t \\
\pro{p'_{\len{s} : \len{t}} = t_{\len{s} : \len{t}}} & \len{t} > \len{s} \cap s = t_{0 : \len{s}} \\
0 & \text{otherwise}
\end{cases}
\end{equation}

Because \(r_j = s\), it is the case that

\begin{equation}
\begin{split}
p'_{\len{s} : \len{t}} &= (r_j \cdot r_{j+1} \cdot r_{j+2} \cdot \ldots)_{\len{s} : \len{t}} \\
&= (r_j \cdot r_{j+1} \cdot r_{j+2} \cdot \ldots)_{\len{r_j} : \len{t}} \\
&= (r_{j+1} \cdot r_{j+2} \cdot \ldots)_{0 : \len{t} - \len{r_j}} \\
&= (r_{j+1} \cdot r_{j+2} \cdot \ldots)_{0 : \len{t} - \len{s}}
\end{split}
\end{equation}

But if \(\pro{r_j = s}\) is the same as \(\pro{r_{j+1} = s}\) for all \(j\) (under the assumption that the distribution of strings on the queue is stationary), then

\begin{equation}
\begin{split}
\pro{p'_{\len{s} : \len{t}} = t_{\len{s} : \len{t}}} &= \pro{(r_{j+1} \cdot r_{j+2} \cdot \ldots)_{0 : \len{t} - \len{s}} = t_{\len{s} : \len{t}}} \\
 &= \pro{(r_j \cdot r_{j+1} \cdot \ldots)_{0 : \len{t} - \len{s}} = t_{\len{s} : \len{t}}} \\
 &= \pro{p'_{0 : \len{t} - \len{s}} = t_{\len{s} : \len{t}}}
\end{split}
\end{equation}

In other words, the probability that the rest of the sequence begins with the rest of \(t\) is the same as the probability that the original sequence begins with the rest of \(t\) (under the assumption of stationarity). This result provides us with a recursive formula for calculating the probability that a string formed by a sequence of concatenated productions begins with a particular string:

\begin{equation}
\begin{split}
\pro{p'_{0 : \len{t}} = t} &= \sum_{s \in \Sigma^*} \pro{r_j = s} \pro{p'_{0 : \len{t}} = t \mid r_j = s} \\
&= \sum_{s \in \Sigma^*} \pro{r_j = s} \begin{cases}
1 & \len{t} \leq \len{s} \cap s_{0 : \len{t}} = t \\
\pro{p'_{\len{s} : \len{t}} = t_{\len{s} : \len{t}}} & \len{t} > \len{s} \cap s = t_{0 : \len{s}} \\
0 & \text{otherwise}
\end{cases} \\
&= \sum_{s \in \Sigma^*} \pro{r_j = s} \begin{cases}
1 & \len{t} \leq \len{s} \cap s_{0 : \len{t}} = t \\
\pro{p'_{0 : \len{t} - \len{s}} = t_{\len{s} : \len{t}}} & \len{t} > \len{s} \cap s = t_{0 : \len{s}} \\
0 & \text{otherwise}
\end{cases}
\end{split}
\end{equation}

In the next subsection, we will present the simplified version of this procedure for a 2-tag system together with an example of its application.

\subsection{Case for a 2-tag system}

Consider the probability that a pair of adjacent symbols randomly selected from \(p\) correspond to \(\pi\), where \(\pi = \pi_0 \pi_1 \in \Sigma^2\). This probability is

\begin{equation}
\begin{split}
\pro{p_i p_{i + 1} = \pi} &= \pro{p_i p_{i + 1} = \pi \cap \bigcup_{s \in \Sigma^*} i \triangleleft s} \\
&= \pro{\bigcup_{s \in \Sigma^*} p_i p_{i + 1} = \pi \cap i \triangleleft s} \\
&= \sum_{s \in \Sigma^*} \pro{p_i p_{i + 1} = \pi \cap i \triangleleft s} \\
&= \sum_{s \in \Sigma^*} \pro{i \triangleleft s} \pro{p_i p_{i + 1} = \pi \mid i \triangleleft s} \\
\end{split}
\end{equation}

In the previous section it was shown that

\begin{equation}
\pro{i \triangleleft s} = \frac{\pro{r_j = s} \len{s}}{\sum_{s' \in \Sigma^*} \pro{r_j = s'} \len{s'}}
\end{equation}

Hence what remains to be determined is \(\pro{p_i p_{i + 1} = \pi \mid i \triangleleft s}\), or the probability of selecting a particular adjacent pair of symbols \(\pi\) given that the first element of that pair belongs to some production equal to \(s\).

The first element of the pair could be found in any position in \(s\). Because \(i\) is distributed randomly and uniformly over \(p\), the position of the selected element is also distributed randomly and uniformly over \(s\).

For example, if \(s = bbb\), then either \(\underline{b}bb\), \(b\underline{b}b\), or \(bb\underline{b}\), where the underline indicates which position has been selected. Since \(i\) is uniformly distributed over every symbol position in \(p\), each symbol position in \(s\) is equally likely to be selected by \(i\), given that \(i \triangleleft s\).

Consider \(s = bbb\) once again. Since every symbol in \(s\) is \(b\), then

\begin{equation}
\pro{p_i = b \mid i \triangleleft bbb} = 1
\end{equation}

What is the probability that \(p_{i+1} = b\)? We know that

\begin{equation}
\pro{p_{i+1} = b \mid \underline{b}bb} = \pro{p_{i+1} = b \mid b\underline{b}b} = 1
\end{equation}

since the symbol following the first and second positions in \(bbb\) is always \(b\). But what about \(\pro{p_{i+1} = b \mid bb\underline{b}}\), namely, the probability that \(b\) follows the \textit{last} symbol in \(s\)? Clearly, in this case, \(p_{i+1}\) will be a symbol belonging to the next production, not the current one.

We can determine \(\pro{p_{i+1} = b \mid bb\underline{b}}\) by considering the \textit{first} symbol of every subsequent production, weighed by the probability of that production.

For example, suppose that

\begin{equation}
\begin{split}
\pro{r_j = bbb} &= 1/4 \\
\pro{r_j = ab} &= 1/4 \\
\pro{r_j = bb} &= 1/4 \\
\pro{r_j = a} &= 1/4
\end{split}
\end{equation}

We then know that

\begin{equation}
\begin{split}
\pro{i \triangleleft bbb} &= 3/8 \\
\pro{i \triangleleft ab} &= 2/8 \\
\pro{i \triangleleft bb} &= 2/8 \\
\pro{i \triangleleft a} &= 1/8
\end{split}
\end{equation}

Hence

\begin{equation}
\begin{split}
\pro{p_{i+1} = b \mid bb\underline{b}}
&= \sum_{s \in \Sigma^*} [s_0 = b] \pro{r_j = s} \\
&= \pro{r_j = bbb} + \pro{r_j = bb} \\
&= 1/4 + 1/4 \\
&= 1/2
\end{split}
\end{equation}

We can then calculate \(\pro{p_{i+1} = b \mid i \triangleleft bbb}\) as

\begin{equation}
\begin{split}
\pro{p_{i+1} = b \mid i \triangleleft bbb}
&= \pro{p_{i+1} = b \cap (\underline{b}bb \cup b\underline{b}b \cup bb\underline{b})} \\
&= 1/3 + 1/3 + 1/3 \cdot \pro{p_{i+1} = b \mid bb\underline{b}} \\
&= 1/3 + 1/3 + 1/6 \\
&= 5/6
\end{split}
\end{equation}

Therefore \(\pro{p_i p_{i + 1} = bb \mid i \triangleleft bbb} = 5/6\). In general,

\begin{equation}
\pro{p_i = \pi_0 \mid i \triangleleft s} = \frac{1}{\lvert s \rvert} \sum_{k \in [0, \len{s})} [s_k = \pi_0]
\end{equation}

and

\begin{equation}
\pro{p_i p_{i + 1} = \pi \mid i \triangleleft s}
= \frac{1}{\lvert s \rvert} \left( [s_{\len{s}-1} = \pi_0] \operatorname{B}_{\pi_1} + \sum_{k \in [0, \len{s} - 1)} [s_k s_{k+1} = \pi] \right)
\end{equation}

where \(\operatorname{B}_{\pi_1}\) is the probability that the next production begins with \(\pi_1\):

\begin{equation}
\begin{split}
\operatorname{B}_\beta
&= \pro{(r_j)_0 = \pi_1} \\
&= \pro{(r_j)_0 = \pi_1 \cap \bigcup_{s' \in \Sigma^*} r_j = s'} \\
&= \pro{\bigcup_{s' \in \Sigma^*} (r_j)_0 = \pi_1 \cap r_j = s'} \\
&= \sum_{s' \in \Sigma^*} \pro{(r_j)_0 = \pi_1 \cap r_j = s'} \\
&= \sum_{s' \in \Sigma^*} \pro{r_j = s'} \pro{(r_j)_0 = \pi_1 \mid r_j = s'} \\
&= \sum_{s' \in \Sigma^*} \pro{r_j = s'} [s'_0 = \pi_1]
\end{split}
\end{equation}

It is possible to determine some properties of the tag system during a particular epoch based on the initial pair probability distribution of that epoch. For example, we can determine the expected queue growth per step by calculating the expected length of the production distribution:

\begin{equation}
\operatorname{E}[\len{r_j}] = \sum_{s' \in \Sigma^*} \pro{r_j = s'} \len{s'}
\end{equation}

and subtracting \(n\) since \(n\) symbols are deleted at each step.

We can also determine the density of different symbols at the beginning of this epoch from the pair distribution by counting the occurrences of those symbols in each production and weighing the counts by the probability of that production being generated, and then summing over all productions.

\subsection{Expected length of the queue}

It is possible to estimate the length of the tag system's queue at any step along its evolution using the methods we have developed.

For an \(n\)-tag system, the duration or number of steps in an epoch is the length of the queue at the beginning of that epoch divided by \(n\), since \(n\) symbols are removed from the front of the queue in each step of the computation.

The expected change in queue length during an epoch is the expected growth per step during that epoch times the number of steps. Hence the expected length of the queue at the beginning of the next epoch is given by

\begin{equation}
\begin{split}
\text{next length} &= \text{current length} + \text{length change} \\
&= \text{current length} + \text{length change per step} \times \text{steps} \\
&= \text{current length} + \text{length change per step} \times \frac{\text{current length}}{n} \\
&= \text{current length} \times \left(1 + \frac{\text{length change per step}}{n}\right) \\
\end{split}
\end{equation}

For example, the predicted queue length at the beginning of every epoch for a particular set of rules, starting with 100 symbols in the queue, is

\begin{center}
\begin{tabular}{ |c|c|c|c|c|c|c|c| } \hline
\multicolumn{8}{|c|}{\( aa \rightarrow aaa,\, ab \rightarrow b,\, ba \rightarrow a,\, bb \rightarrow b \)} \\ \hline
Epoch & 0 & 1 & 2 & 3 & 4 & 5 & 6 \\ \hline\hline
Growth & -0.500 & 0.000 & 0.444 & 0.750 & 0.904 & 0.966 & 0.989 \\ \hline
Length & 100.00 & 75.00 & 75.00 & 91.65 & 126.02 & 182.98 & 271.36 \\ \hline
\end{tabular}
\end{center}

\section{Algorithm}

\subsection{2-tag system simulator}

The following program, written in the Python 3 programming language, simulates a 2-tag system with the set of production rules

\begin{equation}
aa \rightarrow bbb,\, ab \rightarrow ab,\, ba \rightarrow bb,\, bb \rightarrow a
\end{equation}

for 10 epochs, starting with a random configuration of 10 thousand symbols. The program prints the density of each symbol in the queue as well as the length of the queue at the beginning of each epoch:

\begin{lstlisting}
import random

# Returns the production to be appended to the queue
def get_prod(word):
  if word == (0, 0):
    return (1, 1, 1)
  if word == (0, 1):
    return (0, 1)
  if word == (1, 0):
    return (1, 1)
  if word == (1, 1):
    return (0,)

# Creates a queue with a random configuration of specified length
def create_queue(length):
  return tuple(random.choice([0, 1]) for n in range(length))

# Updates the queue by deleting two symbols from the front and
# appending a corresponding production to the back
def update(queue):
  return queue[2:] + get_prod(queue[:2])

# Update the queue until all original symbols have been deleted
def update_epoch(queue):
  for n in range(int(len(queue)/2)):
    queue = update(queue)
  return queue

epochs = 10
queue = create_queue(10000)

for epoch in range(epochs):
  print('Epoch ' + str(epoch))
  print('Length: ' + str(len(queue)))
  print('Density of a symbols: ' + str(queue.count(0)/len(queue)))
  print('Density of b symbols: ' + str(queue.count(1)/len(queue)))
  print('')
  queue = update_epoch(queue)
\end{lstlisting}

An example output for this particular set of rules is

\begin{multicols}{2}
{\tiny
\begin{verbatim}
Epoch 0
Length: 10000
Density of a symbols: 0.5006
Density of b symbols: 0.4994

Epoch 1
Length: 10006
Density of a symbols: 0.24475314811113333
Density of b symbols: 0.7552468518888666

Epoch 2
Length: 7452
Density of a symbols: 0.5017444981213097
Density of b symbols: 0.4982555018786903

Epoch 3
Length: 7465
Density of a symbols: 0.2505023442732753
Density of b symbols: 0.7494976557267247

Epoch 4
Length: 5603
Density of a symbols: 0.5059789398536498
Density of b symbols: 0.49402106014635017

Epoch 5
Length: 5637
Density of a symbols: 0.24942345219088166
Density of b symbols: 0.7505765478091183

Epoch 6
Length: 4224
Density of a symbols: 0.506155303030303
Density of b symbols: 0.49384469696969696

Epoch 7
Length: 4250
Density of a symbols: 0.24776470588235294
Density of b symbols: 0.7522352941176471

Epoch 8
Length: 3178
Density of a symbols: 0.4977973568281938
Density of b symbols: 0.5022026431718062

Epoch 9
Length: 3171
Density of a symbols: 0.25449385052034057
Density of b symbols: 0.7455061494796594
\end{verbatim}
}
\end{multicols}

The following Python code

\begin{figure}[H]
\includegraphics[width=\textwidth]{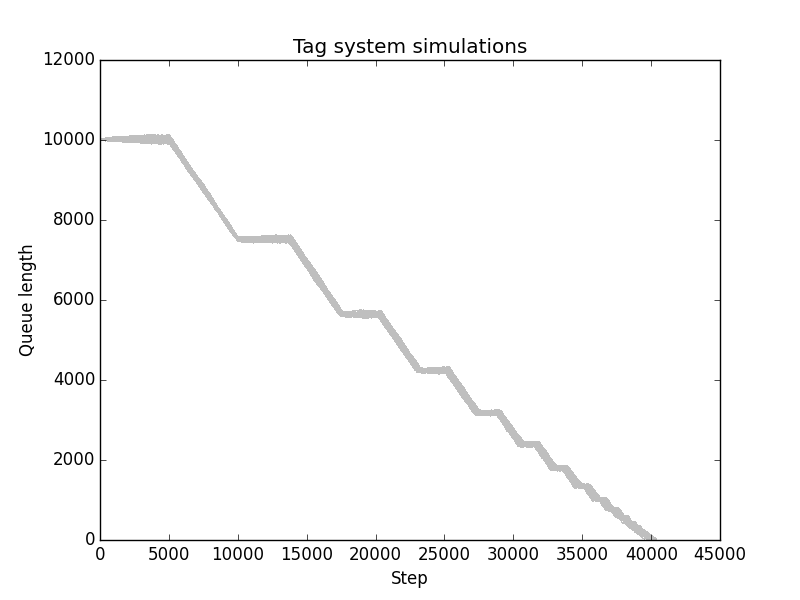}
\caption{}
\label{fig:plot}
\end{figure}

\begin{lstlisting}
import matplotlib.pyplot as plt
plt.title('Tag system simulations')
plt.xlabel('Step')
plt.ylabel('Queue length')
for trial in range(100):
  lengths = []
  queue = create_queue(10000)
  while len(queue) >= 2:
    lengths.append(len(queue))
    queue = update(queue)
  plt.plot(lengths, ',', color='.75')
plt.show()
\end{lstlisting}

plots the queue length at every step for 100 tag systems with the same rule set but with different initial conditions. The plot is shown in figure \ref{fig:plot}.

\subsection{2-tag system predictor}

The following diagram illustrates how the algorithm works: We find the selected pair for each position in every production and, in the case of the last position, consider every second production that could be appended.

\bigskip
\dirtree{%
.1 \(\varepsilon\) \(1\).
    .2 bbb \(\alpha\).
      .3 \underline{bb}b \(\alpha\).
      .3 b\underline{bb} \(\alpha\).
      .3 bb\underline{b} \(\alpha\).
        .4 bb\underline{b b}bb \(\alpha \times 1/4 = \alpha/4\).
        .4 bb\underline{b a}b \(\alpha \times 1/4 = \alpha/4\).
        .4 bb\underline{b b}b \(\alpha \times 1/4 = \alpha/4\).
        .4 bb\underline{b a} \(\alpha \times 1/4 = \alpha/4\).
    .2 ab \(\beta\).
      .3 \underline{ab} \(\beta\).
      .3 a\underline{b} \(\beta\).
        .4 a\underline{b b}bb \(\beta \times 1/4 = \beta/4\).
        .4 a\underline{b a}b \(\beta \times 1/4 = \beta/4\).
        .4 a\underline{b b}b \(\beta \times 1/4 = \beta/4\).
        .4 a\underline{b a} \(\beta \times 1/4 = \beta/4\).
    .2 bb \(\gamma\).
      .3 \underline{bb} \(\gamma\).
      .3 b\underline{b} \(\gamma\).
        .4 b\underline{b b}bb \(\gamma \times 1/4 = \gamma/4\).
        .4 b\underline{b a}b \(\gamma \times 1/4 = \gamma/4\).
        .4 b\underline{b b}b \(\gamma \times 1/4 = \gamma/4\).
        .4 b\underline{b a} \(\gamma \times 1/4 = \gamma/4\).
    .2 a \(\delta\).
      .3 \underline{a} \(\delta\).
        .4 \underline{a b}bb \(\delta \times 1/4 = \delta/4\).
        .4 \underline{a a}b \(\delta \times 1/4 = \delta/4\).
        .4 \underline{a b}b \(\delta \times 1/4 = \delta/4\).
        .4 \underline{a a} \(\delta \times 1/4 = \delta/4\).
}
\bigskip

where \(\alpha + \beta + \gamma + \delta = 1\). Adding up the quantities for each pair yields

\begin{equation}
\begin{split}
&\underline{aa} : \delta/4 + \delta/4 = \delta/2 \\
&\underline{ab} : \beta + \delta/4 + \delta/4 = \beta + \delta/2 \\
&\underline{ba} : \alpha/4 + \alpha/4 + \beta/4 + \beta/4 + \gamma/4 + \gamma/4 = \alpha/2 + \beta/2 + \gamma/2 \\
&\underline{bb} : \alpha + \alpha + \alpha/4 + \alpha/4 + \beta/4 + \beta/4 + \gamma + \gamma/4 + \gamma/4 = 5\alpha/2 + \beta/2 + 3\gamma/2 \\
&\text{total} : \delta/2 + \beta + \delta/2 + \alpha/2 + \beta/2 + \gamma/2 + 5\alpha/2 + \beta/2 + 3\gamma/2 \\
&= 3\alpha + 2\beta + 2\gamma + \delta
\end{split}
\end{equation}

Dividing the quantity of each pair by the total yields

\begin{equation}
\begin{split}
\pro{\underline{aa}} &= \frac{\delta/2}{3\alpha + 2\beta + 2\gamma + \delta} = \frac{\delta}{6\alpha + 4\beta + 4\gamma + 2\delta} \\
\pro{\underline{ab}} &= \frac{\beta + \delta/2}{3\alpha + 2\beta + 2\gamma + \delta} = \frac{2\beta + \delta}{6\alpha + 4\beta + 4\gamma + 2\delta} \\
\pro{\underline{ba}} &= \frac{\alpha/2 + \beta/2 + \gamma/2}{3\alpha + 2\beta + 2\gamma + \delta} = \frac{\alpha + \beta + \gamma}{6\alpha + 4\beta + 4\gamma + 2\delta} \\
\pro{\underline{bb}} &= \frac{5\alpha/2 + \beta/2 + 3\gamma/2}{3\alpha + 2\beta + 2\gamma + \delta} = \frac{5\alpha + \beta + 3\gamma}{6\alpha + 4\beta + 4\gamma + 2\delta}
\end{split}
\end{equation}

For instance, letting \(\alpha = \beta = \gamma = \delta = 1/4\) yields 

\begin{equation}
\begin{split}
6\alpha + 4\beta + 4\gamma + 2\delta = 4 \\
\pro{\underline{aa}} = 1/16 = 0.0625 \\
\pro{\underline{ab}} = 3/16 = 0.1875 \\
\pro{\underline{ba}} = 3/16 = 0.1875 \\
\pro{\underline{bb}} = 9/16 = 0.5625
\end{split}
\end{equation}

which are the probabilities of randomly selecting each pair from the queue, allowing us to predict the productions that will be generated during this epoch. The following program implements this algorithm, using it to predict the large-scale properties of the specified rule set of a 2-tag system:

\begin{lstlisting}
# Returns the production to be appended to the queue
def get_prod(word):
  if word == (0, 0):
    return (1, 1, 1)
  if word == (0, 1):
    return (0, 1)
  if word == (1, 0):
    return (1, 1)
  if word == (1, 1):
    return (0,)

# Returns the production distribution from the word distribution
def get_prod_probs(word_probs, get_prod):
  prod_probs = {}
  for word in word_probs:
    prod = get_prod(word)
    if prod not in prod_probs: prod_probs[prod] = 0
    prod_probs[prod] += word_probs[word]
  return prod_probs

# Normalizes a distribution into a probability distribution
def normalize(probs):
  total = sum(probs.values())
  for event in probs:
    probs[event] /= total
  return probs

# Returns the word distribution from the production distribution
def get_word_probs(prod_probs):
  word_probs = {}
  for prod in prod_probs:
    if len(prod) == 0: continue
    for pos in range(len(prod) - 1):
      word = prod[pos:pos + 2]
      if word not in word_probs: word_probs[word] = 0
      word_probs[word] += prod_probs[prod]
    # Consider every production that could follow the current one
    for prod2 in prod_probs:
      if len(prod2) == 0: continue
      word = (prod[-1], prod2[0])
      if word not in word_probs: word_probs[word] = 0
      word_probs[word] += prod_probs[prod] * prod_probs[prod2]
  return normalize(word_probs)

# Returns the expected density of symbols on the queue
def get_densities(word_probs):
  densities = {}
  for word in word_probs:
    for symbol in word:
      if symbol not in densities: densities[symbol] = 0
      densities[symbol] += word_probs[word]
  return normalize(densities)
      
def get_growth(word_probs, get_prod):
  expected_length = 0
  for word in word_probs:
    expected_length += len(get_prod(word)) * word_probs[word]
  return expected_length - 2

initial_word_probs = {}
initial_word_probs[(0, 0)] = .25
initial_word_probs[(0, 1)] = .25
initial_word_probs[(1, 0)] = .25
initial_word_probs[(1, 1)] = .25

word_probs = {}
prod_probs = {}

word_probs[0] = initial_word_probs
epochs = 10

for epoch in range(epochs):
  prod_probs[epoch + 1] = get_prod_probs(word_probs[epoch], get_prod)
  word_probs[epoch + 1] = get_word_probs(prod_probs[epoch + 1])

length = 10000
for epoch in range(epochs):
  print('Epoch ' + str(epoch))
  densities = get_densities(word_probs[epoch])
  growth = get_growth(word_probs[epoch], get_prod)
  print('Length: ' + str(length))
  print('Density of a symbols: ' + str(densities[0]))
  print('Density of b symbols: ' + str(densities[1]))
  print('')
  length *= (1 + growth/2)
\end{lstlisting}

Notice that the lengths of the string productions being considered inside the \texttt{get\_prod\_probs} function do not appear explicitly because the factors of \(\len{s}\) in the probability of that production being selected from the queue and of selecting a particular position within that production cancel each other.

The output for this particular set of rules is

\begin{multicols}{2}
{\tiny
\begin{verbatim}
Epoch 0
Length: 10000
Density of a symbols: 0.5
Density of b symbols: 0.5

Epoch 1
Length: 10000.0
Density of a symbols: 0.25
Density of b symbols: 0.75

Epoch 2
Length: 7500.0
Density of a symbols: 0.5
Density of b symbols: 0.5

Epoch 3
Length: 7500.0
Density of a symbols: 0.25
Density of b symbols: 0.75

Epoch 4
Length: 5625.0
Density of a symbols: 0.5
Density of b symbols: 0.5

Epoch 5
Length: 5625.0
Density of a symbols: 0.25
Density of b symbols: 0.75

Epoch 6
Length: 4218.75
Density of a symbols: 0.5
Density of b symbols: 0.5

Epoch 7
Length: 4218.75
Density of a symbols: 0.25
Density of b symbols: 0.75

Epoch 8
Length: 3164.0625
Density of a symbols: 0.5
Density of b symbols: 0.5

Epoch 9
Length: 3164.0625
Density of a symbols: 0.25
Density of b symbols: 0.75
\end{verbatim}
}
\end{multicols}

Notice the similarity of these values to those of the output produced by the tag system simulator, and the similarity of the queue lengths that were predicted by the algorithm to those of the sharp changepoints in the plot.

\section{Results}

The following tables show the expected growth of the queue per step along with the density of \(a\) and \(b\) symbols at the beginning of each epoch for different tag system rule sets. These tables also show the expected length of the queue at the beginning of each epoch, assuming an initial reference length of 100.

\begin{center}
\begin{tabular}{ |c|c|c|c|c|c|c|c| } \hline
\multicolumn{8}{|c|}{\( aa \rightarrow aab,\, ab \rightarrow ab,\, ba \rightarrow b,\, bb \rightarrow ba \)} \\ \hline
Epoch & 0 & 1 & 2 & 3 & 4 & 5 & 6 \\ \hline\hline
\(a\) density & 0.500 & 0.500 & 0.467 & 0.455 & 0.450 & 0.449 & 0.448 \\ \hline
\(b\) density & 0.500 & 0.500 & 0.533 & 0.545 & 0.550 & 0.551 & 0.552 \\ \hline
Growth & 0.000 & -0.125 & -0.167 & -0.181 & -0.186 & -0.188 & -0.189 \\ \hline
Length & 100.00 & 100.00 & 93.75 & 85.92 & 78.15 & 70.88 & 64.22 \\ \hline
\end{tabular}
\end{center}

\begin{center}
\begin{tabular}{ |c|c|c|c|c|c|c|c| } \hline
\multicolumn{8}{|c|}{\( aa \rightarrow bb,\, ab \rightarrow bb,\, ba \rightarrow aaa,\, bb \rightarrow bb \)} \\ \hline
Epoch & 0 & 1 & 2 & 3 & 4 & 5 & 6 \\ \hline\hline
\(a\) density & 0.500 & 0.333 & 0.120 & 0.054 & 0.026 & 0.013 & 0.006 \\ \hline
\(b\) density & 0.500 & 0.667 & 0.880 & 0.946 & 0.974 & 0.987 & 0.994 \\ \hline
Growth & 0.250 & 0.083 & 0.037 & 0.017 & 0.008 & 0.004 & 0.002 \\ \hline
Length & 100.00 & 112.50 & 117.17 & 119.34 & 120.35 & 120.83 & 121.07 \\ \hline
\end{tabular}
\end{center}

\begin{center}
\begin{tabular}{ |c|c|c|c|c|c|c|c| } \hline
\multicolumn{8}{|c|}{\( aa \rightarrow bab,\, ab \rightarrow bbb,\, ba \rightarrow aab,\, bb \rightarrow bb \)} \\ \hline
Epoch & 0 & 1 & 2 & 3 & 4 & 5 & 6 \\ \hline\hline
\(a\) density & 0.500 & 0.273 & 0.185 & 0.129 & 0.095 & 0.071 & 0.054 \\ \hline
\(b\) density & 0.500 & 0.727 & 0.815 & 0.871 & 0.905 & 0.929 & 0.946 \\ \hline
Growth & 0.750 & 0.455 & 0.296 & 0.210 & 0.153 & 0.115 & 0.088 \\ \hline
Length & 100.00 & 137.50 & 168.78 & 193.76 & 214.11 & 230.49 & 243.74 \\ \hline
\end{tabular}
\end{center}

\begin{center}
\begin{tabular}{ |c|c|c|c|c|c|c|c| } \hline
\multicolumn{8}{|c|}{\( aa \rightarrow b,\, ab \rightarrow b,\, ba \rightarrow aab,\, bb \rightarrow abb \)} \\ \hline
Epoch & 0 & 1 & 2 & 3 & 4 & 5 & 6 \\ \hline\hline
\(a\) density & 0.500 & 0.375 & 0.389 & 0.400 & 0.398 & 0.397 & 0.397 \\ \hline
\(b\) density & 0.500 & 0.625 & 0.611 & 0.600 & 0.602 & 0.603 & 0.603 \\ \hline
Growth & 0.000 & 0.250 & 0.222 & 0.200 & 0.205 & 0.206 & 0.206 \\ \hline
Length & 100.00 & 100.00 & 112.50 & 124.99 & 137.49 & 151.58 & 167.19 \\ \hline
\end{tabular}
\end{center}

\begin{center}
\begin{tabular}{ |c|c|c|c|c|c|c|c| } \hline
\multicolumn{8}{|c|}{\( aa \rightarrow aa,\, ab \rightarrow ba,\, ba \rightarrow \varepsilon,\, bb \rightarrow ab \)} \\ \hline
Epoch & 0 & 1 & 2 & 3 & 4 & 5 & 6 \\ \hline\hline
\(a\) density & 0.500 & 0.667 & 0.767 & 0.863 & 0.929 & 0.965 & 0.982 \\ \hline
\(b\) density & 0.500 & 0.333 & 0.233 & 0.137 & 0.071 & 0.035 & 0.018 \\ \hline
Growth & -0.500 & -0.571 & -0.500 & -0.306 & -0.152 & -0.073 & -0.036 \\ \hline
Length & 100.00 & 75..00 & 53.59 & 40.19 & 34.04 & 31.45 & 30.31 \\ \hline
\end{tabular}
\end{center}

\begin{center}
\begin{tabular}{ |c|c|c|c|c|c|c|c| } \hline
\multicolumn{8}{|c|}{\( aa \rightarrow bbb,\, ab \rightarrow ab,\, ba \rightarrow bb,\, bb \rightarrow b \)} \\ \hline
Epoch & 0 & 1 & 2 & 3 & 4 & 5 & 6 \\ \hline\hline
\(a\) density & 0.500 & 0.125 & 0.100 & 0.083 & 0.071 & 0.062 & 0.056 \\ \hline
\(b\) density & 0.500 & 0.875 & 0.900 & 0.917 & 0.929 & 0.938 & 0.944 \\ \hline
Growth & 0.000 & -0.750 & -0.800 & -0.833 & -0.857 & -0.875 & -0.889 \\ \hline
Length & 100.00 & 100.00 & 62.50 & 37.50 & 21.88 & 12.51 & 7.034 \\ \hline
\end{tabular}
\end{center}

\begin{center}
\begin{tabular}{ |c|c|c|c|c|c|c|c| } \hline
\multicolumn{8}{|c|}{\( aa \rightarrow aaa,\, ab \rightarrow b,\, ba \rightarrow a,\, bb \rightarrow b \)} \\ \hline
Epoch & 0 & 1 & 2 & 3 & 4 & 5 & 6 \\ \hline\hline
\(a\) density & 0.500 & 0.667 & 0.833 & 0.932 & 0.975 & 0.991 & 0.997 \\ \hline
\(b\) density & 0.500 & 0.333 & 0.167 & 0.068 & 0.025 & 0.009 & 0.003 \\ \hline
Growth & -0.500 & 0.000 & 0.444 & 0.750 & 0.904 & 0.966 & 0.989 \\ \hline
Length & 100.00 & 75.00 & 75.00 & 91.65 & 126.02 & 182.98 & 271.36 \\ \hline
\end{tabular}
\end{center}

\begin{center}
\begin{tabular}{ |c|c|c|c|c|c|c|c| } \hline
\multicolumn{8}{|c|}{\( aa \rightarrow bbb,\, ab \rightarrow ab,\, ba \rightarrow bb,\, bb \rightarrow a \)} \\ \hline
Epoch & 0 & 1 & 2 & 3 & 4 & 5 & 6 \\ \hline\hline
\(a\) density & 0.500 & 0.250 & 0.500 & 0.250 & 0.500 & 0.250 & 0.500 \\ \hline
\(b\) density & 0.500 & 0.750 & 0.500 & 0.750 & 0.500 & 0.750 & 0.500 \\ \hline
Growth & 0.000 & -0.500 & 0.000 & -0.500 & 0.000 & -0.500 & 0.000 \\ \hline
Length & 100.00 & 100.00 & 75.00 & 75.00 & 56.25 & 56.25 & 42.19 \\ \hline
\end{tabular}
\end{center}

The following table compares the values for these properties that were predicted using our methods to values that were obtained from simulations by averaging the results of 10 thousand trials over 6 epochs:

\begin{center}
\begin{tabular}{ |c|c|c|c|c|c|c|c| } \hline
\multicolumn{8}{|c|}{\( aa \rightarrow aaa,\, ab \rightarrow b,\, ba \rightarrow a,\, bb \rightarrow b \)} \\ \hline
Epoch & 0 & 1 & 2 & 3 & 4 & 5 & 6 \\ \hline\hline

\multicolumn{8}{|c|}{Density of \(a\) in queue} \\ \hline
Predicted & 0.500 & 0.667 & 0.833 & 0.932 & 0.975 & 0.991 & 0.997 \\ \hline
Measured & 0.500 & 0.662 & 0.826 & 0.925 & 0.972 & 0.990 & 0.997 \\ \hline
Error & 0.000 & 0.005 & 0.007 & 0.007 & 0.003 & 0.001 & 0.000 \\ \hline\hline

\multicolumn{8}{|c|}{Density of \(b\) in queue} \\ \hline
Predicted & 0.500 & 0.333 & 0.167 & 0.068 & 0.025 & 0.009 & 0.003 \\ \hline
Measured & 0.500 & 0.338 & 0.174 & 0.075 & 0.028 & 0.010 & 0.003 \\ \hline
Error & 0.000 & 0.005 & 0.007 & 0.007 & 0.003 & 0.001 & 0.000 \\ \hline\hline

\multicolumn{8}{|c|}{Length of queue} \\ \hline
Predicted & 100.00 & 75.00 & 75.00 & 91.65 & 126.02 & 182.98 & 271.36 \\ \hline
Measured & 100.00 & 74.98 & 74.98 & 92.24 & 126.80 & 183.74 & 272.16 \\ \hline
Error & 0.00 & 0.02 & 0.02 & 0.59 & 0.78 & 0.76 & 0.80 \\ \hline
\end{tabular}
\end{center}

A similar table can be created for the other rule sets. Other rule sets demonstrate a similar relative error despite the difference in production rules, corroborating the general accuracy and precision of the method.

\section{Conclusion}

In this paper, we have developed a method for predicting the large-scale properties of \(n\)-tag systems directly from their production rules. From the distribution of tuples of \(n\) symbols on the queue at the beginning of an epoch, one can predict the distribution of productions that are generated during that epoch. In turn, from this distribution, one can predict the distribution of tuples of symbols for the next epoch. This process can be repeated for any number of epochs.

From the tuple distribution of an epoch, one can determine various properties of the tag system's evolution within that epoch. For example, one can determine the density of particular symbols on the queue, the growth of the queue, or the length of the queue at any step by linear interpolation between the expected length of the beginning of the current epoch and that of the next epoch.

We have compared the property values predicted using our methods to those measured by performing multiple simulations under random initial configurations. These predictions retain great accuracy even after several epochs.

Investigating other properties of the symbol distribution on the queue, and of the production rules themselves, could yield further insight into the large-scale behavior of tag systems. In particular, we would like to find `shortcuts' for determining whether a particular set of rules creates sharp phase transitions between epochs, as opposed to more gradual changes, or determining whether a particular set of rules create periodic behavior in the large-scale properties of the tag system, as shown in one of the examples. One might also want to determine whether and when the tag system reaches an equilibrium distribution by examining the production rules of the tag system directly.

\nocite{*}

\bibliographystyle{acm}
\bibliography{main}

\end{document}